# IMPROVEMENT OF EMAIL THREATS DETECTION BY USER TRAINING

V.Bernard, P-Y.Cousin, A.Lefaillet, M.Mugaruka, C.Raibaud

Undergraduate Students
ECE Paris School of Engineering
Paris, France

## ABSTRACT

*With the generalization of mobile communication systems, solicitations of all kinds in the form of messages and emails are received by users with increasing proportion of malicious ones. They are customized to pass anti-spam filters and ask the person to click or to open the joined dangerous attachment. Current filters are very inefficient against spear phishing emails. It is proposed to improve the existing filters by taking advantage of user own analysis and feedback to detect all kinds of phishing emails. The method rests upon an interface displaying different warnings to receivers. Analysis shows that users trust their first impression and are often lead to ignore warnings flagged by proposed new interactive system.*

## KEYWORDS

*Phishing, Spam, Human Sensors, User-based Paper*

## 1. INTRODUCTION

The tremendous increase of information traffic on the various available supports has correlatively augmented considerably the flow of useless and even dangerous information, up to the point that, overall, the value of global information content is proportionally decreasing. Mail credibility becomes difficult to discern because of growing useless "information" polluting inboxes. Much worse, there is in parallel a trend difficult to stop, where more and more emails are targeted by computer attacks [1] for stealing information and/or ransoming them, damaging infrastructure or creating botnets. Despite constant surveillance from dedicated watching sites and accelerated up-to-date production of protecting software against viruses and malware, these attacks are now adaptively developed according to the target, the attacker and its final goal [2].

Simplest case corresponds to Phishing with a good-looking email asking user to click or reply [3,4]. Spear-Phishing email is customized to pass anti-spam filters and targeted to a small number of users. Pharming is a Phishing improved version where targeted legitimate site is duplicated in user browser address bar by poisoning the victim DNS server and remapping the target website domain onto pharming site IP address. Least visible and most dangerous Advanced Persistent Threat (APT) attacks are used to extract sensitive data and to create backdoors in user system for later access.

Despite good anti-techniques such as content-based [5-7] and behavior-based [8-11] ones, the attacks are still a serious problem because phishes continually change their ways to defeat the anti-phishing techniques. Also, most existing emails filtering approaches are static easy to defeat by modifying emails content and link strings. To counter Phishing attacks, existing approaches are mostly based on technical tools [12 18], or sometimes in combination with a human sensor [19-20]. Today filters inefficiency against spear-phishing emails constitute a weak point of entrance into supposedly "protected" private emails. In the following the possibility to improve the situation is discussed by implying user participation with adequate training to recognize attacks more easily.

Technical approaches [21-24] attempt to automatically identify fraudulent messages and either discard them immediately or warn the users of their potentially harmful nature. Machine Learning is a good example of technical approach used as a solution to counter phishing attacks [5,25 27]. Test of different algorithms to determine which one has best efficiency gives a value around 90% for all of them [7,28]. However, this is an inappropriate rate given the total volume of phishing attacks. Present approach is different as it combines a technical tool with human interaction, with the intention to determine how much more efficient is user implication for spear phishing detection. Some solutions combining technical tool and human sensor have been tried. For instance, when addon browser BAYESHIELD [29] detects a potential phishing website, it displays a blocking pop-up asking users to use tool Analyzer part in order to detect whether the website is legitimate or not. The software also asks users different questions organized in three categories: a) easy to understand (no complicated terms), b) easy to answer (the questions do not strain their cognitive load), c) educative (users will learn to detect phishing without the help of any tool). The results are interesting as the tool helps users to distinguish legitimate websites from fake ones. Their feedback is very positive as they appreciated the tool support in their decision.

Spear-phishing attacks can be executed on any social media platform. The content of sent messages is usually designed to be interesting for the recipients [3]: it can be based on their hobbies, their personal information, profession, ... Most reviewed studies about phishing count the number of users who filled a form with personal information, credentials or banking information to distinguish users falling into the phishing from users recognizing it. This approach is not pertinent here.

Spear-Phishing containing an APT can infect a system by just clicking on a link or on the attachment. The exploitation is triggered immediately after the user interaction, so it is very important that the user can detect a Spear-Phishing email and do not click on the links inside nor on the attachments. Moreover, if detection rules between Phishing and SpearPhishing differ, they share a common basis but Spear-Phishing is designed to bypass anti-phishing systems and to be very discreet as it is the first step of a big and elaborated attack [4].

The solution proposed in the following is based on the same principle as BAYESHIELD, with the difference that instead of focusing on website detection, attention is focused on email spear-phishing detection and the tool is tested in real conditions.

## 2. SYSTEM DESIGN

### 2.1. The Add-On

Here a THUNDERBIRD (an email application) add-on is developed reacting to a potential threat identification. The addon has been tested on THUNDERBIRD 45.0.x. When it occurs, the receiver must be warned to think about the danger it can be exposed to. For the add-on, a window appears with questions about elements of the received email, like attachment, link, failure in sender identification.

The asked questions are easy to understand. If the email contains a link, the user is warned about its existence. He is asked if the link really redirect towards what it announced it would do and how he can check it. In case the email contains some attachment, the user is asked if he expected to receive a file and is warned about it. The asked questions are easy to understand. If the email contains a link, the user is warned about its existence. He is asked if the link really redirect towards what it announced it would do and how he can check it. In case the email contains some attachment, the user is asked if he expected to receive a file and is warned about it. Finally, in email header, the security SPF Test can assert whether the domain name is real or usurped. If the SPF Fail is not succeeding, which means that it is potentially usurped or is not just recognized, the user is warned that the email may be potentially malicious.

Attachment is detected by looking at attachment part into email header. To detect the link, one just search on the raw text, without html tags, the string "http" or "www". Then the user is supposed to make the appropriate decision. There are different possibilities. First, the email is suspicious and the user discovers it. He reports it as a spam in order to improve the Bayesian filter which decides if an email is dangerous or not. Second, and it is why it is chosen to make a send-email window appear for warning the IT department about the potential danger and letting it take appropriate measures. If the user, despite the warning, still decides to trust the email, it is at his own risk and the warning existence does not change the result. Unfortunately, a legitimate email can make the add-on react. Then, the user can encounter a problem. If the user uses the program for too long, his attention can be reduced. In this case, the user will just throw the emails to spam, like if they were malicious ones. This is probably the most dangerous habit the user can have. But if he keeps his attention focused and takes the time to verify if the emails are legitimate, everything is good.

### 2.2. Mailing

For the mailings, SEES software and POSTFIX on a Kali Linux VIRTUALBOX have been used. POSTFIX has been configured to use one of the domains we bought.

SEES software allows individually sending emails the sender email address, the sender name and the other usual mailing fields of which can be directly set. The software does not allow any individual customization in the body. For logistic reasons, the number of hits per link per population has also been counted. In the experiment performed on site at ECE Paris School of Engineering, the sent emails are not considered as spams by the school webmail (office 365) or by any other email client.

## 3. USER-STUDY METHODOLOGY

The goals of the study are:

1) to compare the click-rate on Phishing and Spear-Phishing mailings between a user group supposedly sensitized to security issues and a nonsensitized one.

2) to compare the click-rate with and without proposed tool.

3) to determine if the tool has helped the users in detecting the junk emails they received.

4) to determine if and how many users identify malicious email.

The developed add-on asks questions to the users on some emails that the email server does not know how to classify: i.e. it is not sure whether the email is legitimate or not.

Thirty participants have been recruited as the testing population to install and use proposed add-on. They were told that the tool was in use to prove that users should be included in junk or phishing detection. Questions have been asked to them on potential dubious points that should help them determine if those emails are illegitimate, in exchange to a small reward for their participation. The control population has been divided into three main groups. The first group counts 86 users who belong to Engineering School. Most of them have basic knowledge on computer security. The second and third groups belong to Business and Management Schools. These users are supposedly not sensitized to computer security. No group was aware of the study. All participants are under 25 years old and undergraduates.

A total of four phishing or spear-phishing emails have been sent so as to have a consistent set of data. The goal of these emails was to have the participants click on a link provided in it. The number of clicks on the links are collected by using a bit.ly account. Each time different links have been used by email: one by school for the control population and one for the testers.

For phases 1 and 2, to make emails look more credible, a domain name similar to chosen "victim" organization has been taken, and an email has been sent from a non-existing person in the company (support@example.com for instance). To make everything look normal, a failure in SMTP protocol, allowing to send an email from any fake or real sender, has been exploited. Protections against this failure exist but must be implemented by the domain you spoof. For legal reasons, permission has been asked to domain's owner or used nonexistent one for the other phases.

Phishing emails differ according to the school for legal reasons. A validation has been asked for each email sent with an usurped identity. Four tests have been conducted corresponding to different users' behavioral aspects.

### 3.1 Test Campaign 1

An email has been sent to all (students) populations informing them that their report is available on the online grade platform. Some of the schools do not use such system but it was anticipated that users would click anyway without checking for basic signs of phishing (wrong URL, unknown sender, breaking the habit) because grades are a highly sensitive subject in this period and many students fear failing their courses. The email was sent from an unknown sender (who sounds like a real staff working there) with a fake email address.

### 3.2 Test Campaign 2

Here, the email differs depending on the targeted school. For the first one, it was an email about a change in their timetable. For the other ones, it was an email informing the user that a tutorial about the use of the printers were available. It was sent from a legitimate address.

### 3.3 Test Campaign 3

For third phase, a commercial ad was sent by email. Since the first semester just ended, the "shop" was offering 20% off the alcohol. The email invited users to click on a link to discover the shops participating to the operation. The clues for the users to determine that it was not legitimate were: the obfuscated link, no possibility to unsubscribe the newsletter (it is mandatory in the EU), the lack of any shop with this name and finally, that they did not subscribe to any information of this kind.

### 3.4 Test Campaign 4

In last test, an email was sent informing the students they had submitted their work on the school moodle. Of course, they did not. The course did not even exist. The sender address was unrelated to the school but the expedient name sounds legitimate, i.e. its name is "Name of the school" and its address this@isaspam.org . This phase allows verify what the user checks when they receive shady emails.

## 4. RESULTS

### 4.1 Global Results

The population sizes are: 86 for School 1 (S1- Engineering School), 60 for School 2 (S2- Business and Management School) and 220 for School 3 (S3- Business and Management School). To analyze the results, the following "global click response" (GCR) coefficient $\gamma$ given by (1) will be used

$$\gamma = 10^2\, \chi.\nu \tag{1}$$

where $\chi$ = {Click Number/Number of Clicking Persons} and $\nu$ = {Number of Clicking Persons/Tested Population Number}. This coefficient is grouping two different elements {the number of clickers in a population, the number of clicks per clicking person} together, depending respectively on tested population number and individual response of tested groups. Even if $\chi$ and $\nu$ are generally independently required for specific study, GCR $\gamma$ is for present discussion representative of tested population response. The fact that $\gamma > 10^2$ because $\chi > 1$ is here useful as it enhances the difference between the different tested groups. The most successful tests are first and last ones. The first test corresponds to the most advanced attack and the hardest to detect. The last one was testing user habits and awareness level. There is no data for S2 because authorization was not obtained from school administration. S1 students clicked the less on proposed links. During first test campaign, a huge spike for S2 students reaches around 360%, see Fig. 1.

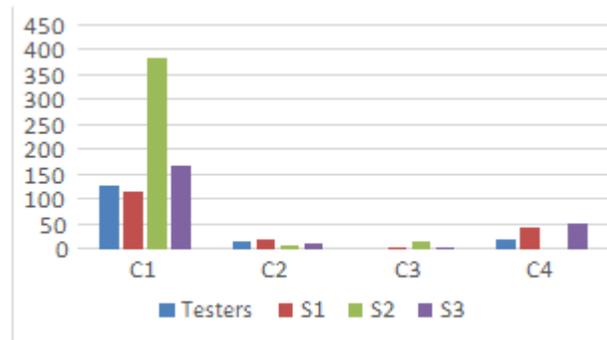

Figure 1: GCR λ per Population per Campaign Cj

S3 students clicked about 170% and S1 students around 110%. Testers performed worse, about 130%. For tests 2 and 3, results are quite similar for every population. For test 2, the average click rate is around 15% and for test 3 about 5%, except for S2, with a click rate of 14%. None of the testers clicked on this mail. For the last test campaign, the click rate is about 44% for S1, 50% for S3 and 21% for volunteer population.

Relevance of proposed tool can be analyzed on Fig. 2 It compares the ratio clicks/control population with the clicks/test population. It brings to light that the curves are quite similar and follow the same tendencies. One can distinguish for test campaign 4 a small difference. Control population clicked around 40% whereas testers did 20%.

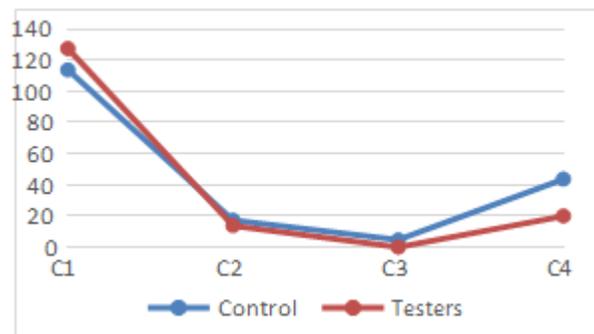

Figure 2: Evolution of GCR λ for Control and Recruited Population.

## 4.2 Results for Test Campaigns 1 and 4

As campaigns 2 and 3 had a very low click rate, the daily analysis was dismissed. The percentages presented in this section are calculated as: $\frac{Number of clicks on day X}{Total number of click} * 100$    (2)

### 4.2.1 First Campaign

During this campaign, the main objective was to create an email the users would crave to read and click on associated link. As it was the end of the semester, the created email was announcing that the results were available on the grade platform. A particular attention has been taken to every detail, name, template, etc. The fake email was based on a similar email sent by the administration of the schools two years before.

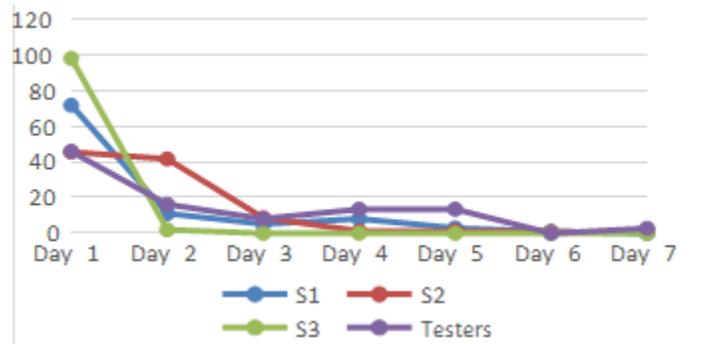

Figure 3: First Test GCR λ per Day.

There was no similar email sent since then. Fig. 3 shows that S3 massively click on the link (around 160%) the first day, but not after. About 70% of S1 clicks are the first day. There is about 10% of S1 clicks on day 2, 3 and 4. S2 and tester clicks are around 45% the first day and stabilize for S2 the second day and quickly drop to zero during day 3 and 4. For testers, click percentage stabilizes around 15% until day 6 and then drops to around 3%.

**4.2.2 Fourth Campaign**

For last test, a new phishing email was sent but this time, with more intriguing features. The designed was basic and the sender address had nothing to do with something remotely legitimate. If the user read the email address, he directly understands the email is a setup. The goal was to determine if users check for basic signs of spam in an email.

Fig. 4 shows that, S1 click on the link the first day but not after. About 80% of tester clicks are the first day. There is about 20% tester clicks on day 2 and none after. S3's clicks are about 60% the first day, it quickly stabilizes around 15% on day 2 and 3. It drops to roughly 10% on day 4.

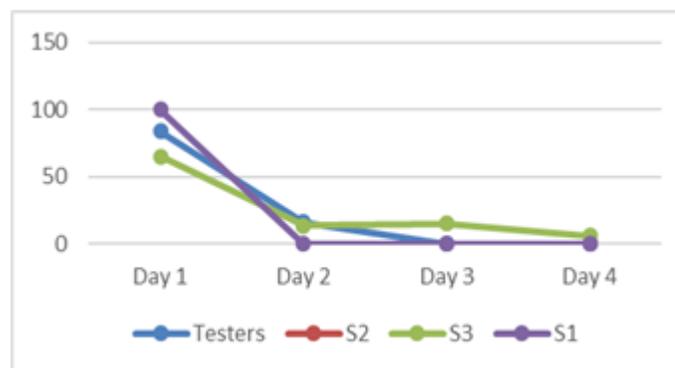

Figure 4: Fourth Test GCR λ per Day

## 5. ANALYSIS

To validate them, survey covering all previously sent emails was sent to all participants asking why they trusted the emails or not, or why they thought it was legitimate.

1)- It cannot be as certain from reported results that there is no statistically convincing data confirming the improvement of phishing detection with proposed tool.

2)- Many users rely on the pair {sender name/subject} to determine whether the email they received is legitimate or not. This is further reinforced by the fourth test phase results. The sender email address was off. It did not even look like something legitimate. However, one still observed a 40~50% click rate on the links. To further check this hypothesis, the email sent was about a homework the students submitted online, but in a subject outside their cursus, after it was checked they did not submit a homework in any subject at all recently. This result can be explained by the fact that whether on a PC or a mobile phone, it is the two first information available. Furthermore, on a mobile, they are generally the only one available, the email address being usually hidden.

3)- There is no real discrepancy between supposedly sensitized and non-sensitized populations ones. They all clicked in similar proportions during the different test campaigns (difference <5%). It can be seen that test population did not click on the links at all during third test whereas the other populations had non-null click rate.

In first test campaign, the click/population number ratio is above 1. The result is explained by users' frustration not to get the promised information and clicking again in believing it was an error. The link redirected them to the school extranet that they, subconsciously, consider as safe and legitimate. Others clicked in the following days thinking the information would be posted later. It is possible that the greater is the ratio, the more difficulty it indicates for the "victims" to spot that it was a spear-phishing.

4)-The high score of first test campaign comes from arousal of students' curiosity and from their need for reassurance against a possible catch-up exam. Therefore, when seeing the email, they thought it was legitimate and did not further check for a phishing. Even those with the add-on did not pay attention to the alert.

The spear-phishing was a high quality one: the message, subject, sending timing and sender were conceived to catch the receiver. A technical fault was exploited to use the domain name of their school and be shown as legitimate. Results are similar to previous studies with trained users [30]
The family and given names of sender person were like a real staff member. Students' feedback shows that they thought they knew the created person.

5)-This is further reinforced in the third test during which few clicks are counted: 4% for S3, 7% for S2, 4% for S1, and 0% for recruited population. It was sent like a normal phishing would be, which can explain their improved awareness. Also, the subject might not be of interest for everyone (alcohol price reduction).

The same can be said for the second test (email about their timetables and printers), the emails had an informative objective and did not present any priority need. Many users said that they deleted it without looking at it like they usually do for this kind of email. When asked about its legitimacy, they answered that it seemed legitimate but it did not interest them. Therefore, they made a mistake in their analysis but it was a benign one because email deletion protected them from possible harm.

6)- Furthermore, many users have bad IT habits. They click on the links present in the emails instead of accessing the website by their own means. They do not pay attention to details (sender email, mistakes, sending legitimacy). When comparing the number of testers to the amount of feedbacks received from the add-on, it appears that users did not feel the need to use proposed extension because they did not see problems in the emails or just did not pay attention to pop-up [23,24]. When comparing present study to other ones, it was supposed that the pop-up did not block the user enough and it could easily be closed. For technical reasons, the pop-up was only monochrome which might not have attracted enough user attention. Also its repetitive aspect probably contributed to the low amount of feedback from the add-on.

## 6. CONCLUSIONS

When compared to other similar ones, present study has been conducted in life conditions and not in artificial environment. Though more realistic, this might have had a negative impact on present results. The objective of the study was to set up a technical solution involving user interaction in spam and phishing more effective detection. The intention was to demonstrate that user inclusion in the fight (wrestling) against Phishing was more effective than just with the machine. After several test campaigns, two observations can be highlighted. First, within the studied framework, the results show that proposed tool does not always improve Phishing detection. Aside inherent limitations of proposed method, the homogeneous education level of test populations could introduce some bias in the results. For instance, some related to many users' bad IT habits, when they base themselves on the combination {sender name/subject} to determine if received email is justifiable or not, without paying attention to elements typically present in impish messages, etc. This would suggest that there is no "universal" best response for user-technical tool collaboration which is depending on both mail subject and warnings nature and ergonomic presentation in tool display. This population sensitivity problem will be considered elsewhere. In particular circumstances, the use of purely technical solution is more efficient than hybrid one including user interaction will be analyzed, because they correspond to situations where user risks to deliver bad information to the tool (for example for messages considered as phishing ones by the tool but considered as legitimate by the user).

**ACKNOWLEDGEMENTS**

The authors are very much indebted to ECE Paris School of Engineering for having provided the necessary setup within which the study has been developed, to Drs M. Kadiri and V. Nuzzo for advices and guidance during the project, Antoine Joly for his proof-reading, and to Pr. M. Cotsaftis for help in preparation of the manuscript.

**REFERENCES**

[1]   The number of detected phishing attacks has increased from 170000 in 2005 to 440000 in 2009

[2]   D. Gudkova, M. Vergelis, N. Demidova : Spam and Phishing in Q3 Securelist, 2016. Spammers and fraudsters are now more interested to make their email contents look as normal as possible, as if they believed a significant proportion of users have mastered the basics of Internet security and can spot a fake threat

[3]   J. Hong : The Current State of Phishing Attacks, Comm.of the ACM, 55(1), pp.74–81, 2012

[4]   L. James : Phishing Exposed. Syngress, 2005

[5]   I. Fette, N. Sadeh, A. Tomasic : Learning to Detect Phishing Emails, Proc. 16th Intern. Conf. on World Wide Web (WWW'07), Alberta, Canada, pp.649–656, ACM, 2007

[6]   M. Chandrasekaran, K. Narayanan, S. Upadhyaya : Phishing Email Detection Based on Structural Properties, NYS Cyber Security Conference, 2006

[7]   S. Abu-Nimeh, D. Nappa, X. Wang, S. Nair : A Comparison of Machine Learning Techniques for Phishing Detection, Proc. Anti-phishing Working Groups, 2nd Annual ECRIME Researchers Summit, pp.60–69, New York, NY, USA, ACM, 2007.

[8]   J. Zhang, Z. Du, W. Liu : A Behaviour-based Detection Approach to Mass-Mailing Host, Proc.6th Intern. Conf. on Machine Learning and Cybernetics, Vol.4, pp.2140-2144, 2007

[9]   F. Toolan, J. Carthy : Feature Selection for Spam and Phishing Detection, Proc. ECRIME Researchers Summit, pp.1-12, 2010

[10]  N.A. Syed, N. Feamster, A. Gray : Learning To Predict Bad Behaviour, NIPS 2007 Workshop on Machine Learning in Adversarial Environments for Computer Security, 2008.

[11]  I.R.A. Hamid, J. Abawajy, TH. Kim : Using Feature Selection and Classification Scheme for Automating Phishing Email Detection, Studies in Informatics and Control, Vol.22 (1), pp.61-70, 2013

[12]  C. Ludl, S. McAllister, E. Kirda, C. Kruegel : On the Effectiveness of Techniques to Detect Phishing Sites, Detection of Intrusions and Malware, and Vulnerability Assessment, pp.20–39, Springer, 2007

[13]  Y. Zhang, J.I. Hong, L.F. Cranor : CANTINA: a Content-based Approach to Detecting Phishing Web Sites, Proc. 16th International Conference on World Wide Web (WWW'07), Alberta, Canada, pp.639– 648, ACM, May 2007

[14]  M. Wu : Fighting Phishing at the User Interface, PhD Thesis, Computer Science and Engineering, MIT, 2006

[15]  S. Garera, N. Provos, M. Chew, A.D. Rubin : A Framework for Detection and Measurement of Phishing Attacks, Proc. 5th ACM Workshop on Recurring Malcode, WORM 07, ACM, New York, NY," USA, pp.1-8, 2007

[16]  C. Whittaker, B. Ryner, M. Nazif : Large Scale Automatic Classification of Phishing Pages, Proc. 17th Annual Network and Distributed System Security Symposium, NDSS 10, San Diego, CA, USA, 2010 "

[17]  R.B. Basnet, A.H. Sung, Q. Liu : Rule Based Phishing Attack Detection, Proc. Int. Conf. Security and Management, SAM11, Las Vegas, NV, USA, 2011

[18]  M. Cova, C. Kuregel, G. Vigna : Detection and Analysis of Drive-bydownload Attacks and Malicious JavaScript Code, Proc. Intern. World Wide Web Conference (WWW'10), Rayleigh, North Carolina, USA, pp.281–290. ACM, 2010

[19]  B.K. Wiederhold : The Role of Psychology in Enhancing Cybersecurity, Cyberpsychology, Behavior, and Social Networking, Vol.17, pp.131– 132, 2014

Computer Science & Information Technology (CS & IT)

[20] R.W. Proctor , Jin Chen : The Role of Human Factors Ergonomics in the Science of Security, Human Factors, Vol.57(5),  2015

[21] J.S. Downs, D. Barbagallo, A. Acquisti : Predictors of Risky Decisions: Improving Judgment and Decision Making Based on Evidence from Phishing Attacks, Neuro-Economics, Judgment, and Decision Making, E.A. Wilhelms, V.F. Reyna, eds., pp.239–253, New York, NY: Psychology Press, 2015

[22] X. Dong, J. Clark, J. Jacob : User Behavior Based Phishing Websites Detection, Proc. 2008 Intern. Multi-conf. on Computer Science and Information Technology (IMCSIT'08), Wisla, Poland, pp.783–790, IEEE, 2008.

[23] L.F. Cranor, S. Egelman, J. Hong, Y. Zhang : Phinding phish: An Evaluation of Anti-phishing Toolbars, Techn. Rept CMU-CyLab-06018, CMU, November 2006.

[24] M. Wu, R.C. Miller, S.L. Garfinkel : Do Security Toolbar Actually Prevent Phishing Attacks, Proc. SIGCHI Conf. on Human Factors in Computing Systems, pp.601-610, ACM, 2006

[25] R. Basnet, S. Mukkamala, A. Sung : Detection of Phishing Attacks: A Machine Learning, Soft Computing Applications in Industry, Studies in Fuzziness and Soft Computing, B. Parsad, ed., Vol.226, pp.373–383. Springer, 2008

[26] V. Dutt, Y.S. Ahn, C. Gonzalez : Cyber Situation Awareness: Modeling Detection of Cyber Attacks with Instance-based Learning Theory, Human Factors, Vol.55, pp.605–618, 2013

[27] R. Basnet, A. Sung, Q. Liu : Learning to Detect Phishing URLs, Intern.J. Research in Engineering and Technology (IJRET), Vol.3(6), pp.11– 24, 2014

[28] G. Sharma, A. Tiwari : A Review of Phishing URL Detection using Machine Learning Systems, Intern. J. Digital Application and Contemporary Research (IJDACR), Vol.4(2), Sept. 2015

[29] P. Likarish, D. Dunbar, J. Hourcade et al.: BAYESHIELD: Conversational Anti-Phishing User Interface, SOUPS : Proc. 5th Symp. on Usable Privacy and Security, 2009.

[30] R.C. Dodge, C. Carve, A.J. Fergusson : Phishing for User Security  Awareness, Computers and Security, Vol.26(1), pp.73-80, 2007